\documentclass[twocolumn,showpacs,preprintnumbers,amsmath,amssymb]{revtex4}
\usepackage{epsfig}
\usepackage{bbm}

\newcommand {\he}{$^4${He} }

\newcommand {\dimer}{$^4${He}$_2$ }

\newcommand {\be}{\begin{equation}}
\newcommand {\ee}{\end{equation}}
\newcommand {\bea}{\begin{eqnarray}}
\newcommand {\ea}{\end{eqnarray*}}
\newcommand {\ba}{\begin{eqnarray*}}
\newcommand {\eea}{\end{eqnarray}}

\newcommand {\ham} {{\mathcal H}}

\newcommand {\bra}{\langle}
\newcommand {\ket}{\rangle}

\newcommand {\refeq}[1] {(\ref{#1})}
\newcommand{\bm}[1]{ \mbox{\boldmath $#1$}  }

\begin{document}
\title{General integral relations for the description of scattering states using the hyperspherical adiabatic basis}

\author{C. Romero-Redondo and E. Garrido}
\affiliation{Instituto de Estructura de la Materia, CSIC, Serrano 123, E-28006 Madrid, Spain}
\author{P. Barletta}
\affiliation{Department of Physics and Astronomy, University College London, 
Gower Street, London WC1E 6BT, United Kingdom}
\author{A. Kievsky and M. Viviani}
\affiliation{Istituto Nazionale di Fisica Nucleare, Largo Pontecorvo 3, 56100 Pisa, Italy}

\begin{abstract}
In this work we investigate 1+2 reactions within the framework of the hyperspherical adiabatic
expansion method. To this aim two integral relations, derived from the Kohn variational principle,
are used. A detailed derivation of these relations is shown. The expressions derived are general,
not restricted to relative $s$ partial waves, and with applicability in multichannel reactions.
The convergence of the ${\cal K}$-matrix in terms of the adiabatic potentials is investigated. 
Together with a simple model case used as a test for the method, we show results for the collision 
of a $^4$He atom on a \dimer dimer (only the elastic channel open), and for collisions involving a 
$^6$Li and two $^4$He atoms (two channels open).
\end{abstract}


\maketitle

\section{Introduction}

Calculation of phase shifts (or the ${\cal K}$-matrix) for a given reaction is often complicated
by the necessity of knowing the wave function of the full system at large distances.
Extraction of the phase shifts can be in principle achieved by comparison of the large distance 
part of the wave
function with its known analytic asymptotic expression. For processes involving only two particles
(1+1 collisions) this procedure can be easily implemented, and therefore the phase shifts can be
computed. However, the more particles involved in the reaction the more difficult the calculation of an
accurate wave function at large distances, or at least the more expensive from the computational
point of view. Therefore, when increasing the number of particles the extraction of the phase shifts
becomes progressively more and more complicated. In nuclear physics, collisions involving three
and four nucleons have been extensively studied
solving the Faddeev ($A=3$) and Faddeev-Yakubovsky ($A=4$)
equations~\cite{gloeckle94,deltuva07}, and the Hyperspherical Harmonic (HH)
expansion in conjunction with the Kohn Variational Principle (KVP)~\cite{kiev08,kiev01}.
These methods have been tested through different benchmarks~\cite{benchmark1,benchmark2}.
When the interaction between the particles presents a hard core, as in the case of the atom-atom
interaction, a direct application of these techniques could be problematic. 
The Faddeev equations has been modified to deal with a hard core
repulsion~\cite{motovilov} and, in the case of the HH expansion, a correlation factor
has been included~\cite{bar01}. In addition the Hyperspherical Adiabatic (HA)
expansion method has proven to be a very efficient tool~\cite{nie01}.

In the case of atom-atom interactions, the HA expansion shows a particularly fast range
of convergence in the description of bound states, as has been shown for example in
Ref.~\cite{blume00} for the description of rare gas trimers. In the past years there was
a systematic use of the HA expansion in the description of three-atom systems in the
ultracold regime (see for example Refs.\cite{esry2008,greene2010} and references therein). 
These applications rise the question about
the convergence properties of the HA method for scattering states, in particular in
the description of a $1+2$ collision. In principle the HA expansion could be applied
to describe such a process since it leads to a clean distinction between all the possible 
incoming and outgoing channels. However, as was recently showed, the convergence of
the expansion slows down significantly in applications directed to describe low energy scattering 
states~\cite{bar09b}. This problem appears at the moment of applying the proper boundary
conditions to the hyperradial functions. In fact, in the HA expansion, the hyperradial
functions are obtained solving an infinite system of equations in the hyperradial variable $\rho$ 
and the convergence of the expansion is studied by increasing the number of equations
considered after truncation of the system. For describing a $1+2$ collision, the hyperradial
functions are obtained requiring an hyperradial plane wave behavior as $\rho\rightarrow\infty$.
However, in such a process, the plane wave behavior results in the relative distance
between the incident particle and the center of mass of the two-body bound system. The
equivalence between both descriptions happens at $\rho\approx\infty$ or, in other words,
by including a very large number of hyperradial functions in the solutions. This is the 
cause of the extremely slow observed convergence.

In Ref.\cite{bar09} the authors introduced a general method to compute the phase shift
from two integral relations that involve only the internal part of the wave function. 
This method is a generalization to more than two particles of the integral relations 
given in \cite{har67,hol72} and it is derived from the KVP. In the case of the HA
expansion, in Ref.~\cite{bar09} was shown that for a 1+2 reactions, the use of the 
integral relations allows to determine the phase-shift with a pattern of convergence 
similar to a bound state calculation. Therefore, thanks to the integral relations, the 
hyperspherical adiabatic expansion method appears
as a powerful tool also to describe scattering processes.

The purpose of this work is to show in details the use of the integral relations in conjunction
with the HA expansion method to describe scattering states. In Ref.~\cite{bar09}
the particular case of a 1+2 reaction with only the elastic channel open, and with only relative
$s$-waves involved, was considered. The applicability of the method is not limited
to this particular case. In this work we shall consider processes involving $\ell\geq 0$ relative angular
momenta, and we shall derive the integral relations for the general case in which more than one channel is open.
The only limitation is that we shall restrict ourselves to energies below the breakup threshold.
Above it infinitely many adiabatic terms are in principle needed to describe the breakup channel,
and although the same procedure could be used to describe it, we leave this particular case for a more
careful investigation in a forthcoming work.

A different aspect is the applicability of the method to describe 1+$N$ reactions with $A=1+N>3$.
In this case, the main difficulty is to obtain the $N$+1 wave function in the internal region and
the $N$-body bound state function describing the asymptotic configuration. With this information, the integral
relations apply exactly the same way as for the 1+2 case, but replacing the bound dimer wave function by
the corresponding bound $N$-body wave function. The extension of the adiabatic expansion to
describe more than three-particles is possible. The dependence of the hyperangular part consists in
$(3N-4)$ hyperangles and, in the case of systems of identical particles, the problem of
constructing a $A$-body wave function with the proper statistic has to be faced. First applications
of the HA expansion to describe a four-body system already appeared~\cite{wang09}.
In this work, however, we restrict the discussion to $1+2$ reactions.

In section II we describe the details of the formalism, first describing the adiabatic expansion
in a multichannel reaction, and second showing how the corresponding ${\cal K}$-matrix (or equivalently the
${\cal S}$-matrix) can be obtained from the asymptotic wave function. In section III the integral relations
for the same multichannel reaction are derived. They permit to extract the ${\cal K}$- (or ${\cal S}$-) matrix
requiring only knowledge of the internal part of the wave function. The results are shown in section IV.
In section \ref{sec2a} we consider a test case with only the elastic channel open. We investigate a three-body process 
which is fully equivalent to a two-body reaction, for which the phase shifts can be easily
computed. This can then be used to test the
accuracy of the integral relations method as well as the convergence pattern in the adiabatic expansion
when $\ell > 0$ partial waves are involved. In section IV B we investigate
the elastic collision between a $^4$He atom and the weakly bound $(^4$He$)_2$ dimer. 
Finally, in IV C we apply the method to the collision involving a $^6$Li and two $^4$He atoms. 
In particular we shall consider incident energies such that the two possible incoming and 
outgoing channels, $(^4$He,$(^4$He$-^6$Li$))$ and $(^6$Li,$(^4$He$)_2)$ are both open. 
The summary and the conclusions are
given in section V. In appendix A we show the derivation of the Kohn 
Variational Principle for a multichannel process and, finally, in appendix B
 we have collected some technical details of the use of the integral
relations when projected two-body potentials are employed.

\section{Formalism}

\subsection{General features of the HA expansion}

In this work we consider a process where a particle hits a bound two-body system. We assume
the incident energy to be below the breakup threshold in three particles. 
This means that the total three-body
energy $E$, which is the sum of the incident energy $k^2/2\mu>0$ ($\mu$ being the reduced mass between
the incident particle and the dimer) and the two-body binding energy 
$E_{2b}$, is negative. In this way only elastic, inelastic, 
and rearrangement processes are possible.

The reaction under study is therefore a three-body process, which as usual, can be described
through the $\bm{x}$ and $\bm{y}$ Jacobi coordinates: 
\bea
 & & \bm{x}_i= \sqrt{\frac{m_j m_k}{m(m_j+m_k)}}(\bm{r}_j-\bm{r}_k)   \\
 & & \bm{y}_i= \sqrt{\frac{m_i(m_j+m_k)}{m(m_i+m_j+m_k)}}
        \left(\bm{r}_i-\frac{m_j\bm{r}_j+m_k \bm{r}_k}{m_j+m_k}\right) \nonumber
\eea
where $m_i$ and $\bm{r}_i$ are the mass and coordinate of particle $i$ and 
$m$ is an arbitrary normalization mass.
 From the Jacobi coordinates one can construct
the hyperspherical coordinates, which contain a radial one, the so-called hyperradius $\rho$ 
($\rho^2=\sqrt{x^2_i+y^2_i}$) and the five hyperangles $\Omega$ 
($[\Omega]\equiv [\alpha_i,\Omega_x,\Omega_y]$).
The hyperangle $\alpha_i$ is defined as $\tan\alpha_i=x_i/y_i$ and $\Omega_x$ and $\Omega_y$ give the directions
of $\bm{x}_i$ and $\bm{y}_i$. The five hyperangles depend on the particular ordering of the particles chosen
in the definition of the Jacobi variables. Three different sets are possible by cyclic
permutations of the indexes $i,j,k$. In the following the Jacobi coordinates $\bm{x}$ and $\bm{y}$
and the corresponding hyperangular coordinates are given using the natural ordering of the
particles $i,j,k\equiv 1,2,3$.

Following Ref.~\cite{nie01} we give a brief description of the HA method.
In hyperspherical coordinates the Hamiltonian operator $\hat{\ham}$
takes the form:
\be
\hat{\ham} =  -\frac{\hbar^2}{2 m} \hat{T}_\rho + \frac{\hbar^2}{2 m \rho^2}\hat{G}^2
+ V(\rho,\Omega)
 =  -\frac{\hbar^2}{2 m} \hat{T}_\rho + \hat{{\cal H}}_\Omega    ,
\label{eq1}
\ee
where $\hat{T}_\rho=\frac{\partial^2}{\partial\rho^2}+\frac{5}{\rho}\frac{\partial}{\partial\rho}$ 
is the hyperradial kinetic energy operator, $\hat{G}^2$ is the grand-angular
operator and $V(\rho,\Omega)=\sum_i V_i(x_i)$ is the potential energy ($i$ runs over the 
three Jacobi systems). 

The adiabatic expansion is based on the assumption that when describing a particular process,
the hyperangles vary much faster than the hyperradius $\rho$. Under this assumption
it is possible to solve the Schr\"{o}dinger equation $(\hat{\cal H}-E)\Psi=0$ in two steps. 
In the first one the angular part is solved for a set of fixed values of $\rho$. This amounts 
to solve the eigenvalue problem 
\be
\hat{{\cal H}}_\Omega \Phi_n(\rho,\Omega)=\frac{\hbar^2}{2 m} \frac{1}{\rho^2}
\lambda_n(\rho) \Phi_n(\rho,\Omega)
\label{eq2}
\ee
for each $\rho$, which is treated as a parameter. 

The angular functions $\{\Phi_n(\rho, \Omega)\}$ are used to construct the HA
basis in which the basis elements form an orthonormal basis for each value of $\rho$.
The full three-body wave function is then expanded as:
\be
\Psi(\bm{x},\bm{y}) = \frac{1}{\rho^{5/2}}\sum_{n=1}^\infty f_n(\rho) \Phi_n(\rho,\Omega).
\label{eq3}
\ee
Obviously the summation above has to be truncated, and only a finite number $n_A$ of adiabatic terms are
included in the calculation. For simplicity we are omitting in $\Psi$, $f_n$, and $\Phi_n$ the quantum
numbers giving the total three-body angular momentum and its projection.

In a second step, the radial wave functions $f_n(\rho)$ in the expansion of Eq.(\ref{eq3}) are obtained
after solving the following coupled set of radial equations:
\be
\sum_{n'=1}^{n_A}\left(\hat{\cal H}_{nn'}-E \delta_{nn'}\right) f_{n'}(\rho)=0,
\label{eq4}
\ee
where the operator $\hat{\cal H}_{nn'}$ acts on the radial functions and takes the form
\begin{equation}
\hat{\cal H}_{nn}(\rho)=\frac{\hbar^2}{2m}
\left[ -\frac{d^2}{d\rho^2} -Q_{nn}(\rho) + \frac{1}{\rho^2}
\left( \lambda_n(\rho)+\frac{15}{4} \right) \right]
\label{eq5}
\end{equation}
for the diagonal terms, and
\begin{equation}
\hat{\cal H}_{nn'}= -\frac{\hbar^2}{2m}
\left( 2 P_{n n'}(\rho) \frac{d}{d\rho} + Q_{n n'}(\rho) \right)
\label{eq6}
\end{equation}
when $n\neq n'$.

The coupling terms $P_{nn'}$ and $Q_{nn'}$ in the expressions above follow from the dependence on $\rho$ 
of the HA basis. Their explicit form is
\bea
& & P_{n n'}(\rho)=\Big\bra \Phi_n(\rho,\Omega) \Big|\frac{\partial}{\partial \rho} \Big| 
                           \Phi_{n^\prime}(\rho,\Omega) \Big\ket_\Omega  \nonumber \\
& & Q_{n n'}(\rho)=\Big\bra \Phi_n(\rho,\Omega) \Big|\frac{\partial^2}{\partial \rho^2} \Big| 
                           \Phi_{n^\prime}(\rho,\Omega) \Big\ket_\Omega, 
\label{coup}
\eea
where $\langle \rangle_\Omega$ represents integration over the five hyperangles only.

The one-dimensional set of coupled differential equations given in Eq.(\ref{eq4}) 
can be written in a matrix form as:
\be
\left(
	\begin{array}{cccc}
	\hat{\cal H}_{11}-E & \hat{\cal H}_{12} & \cdots & \hat{\cal H}_{1n_A} \\
	\hat{\cal H}_{21} & \hat{\cal H}_{22}-E & \cdots & \hat{\cal H}_{2n_A} \\
	 \vdots       & \vdots        & \vdots & \vdots        \\
	\hat{\cal H}_{n_A1} & \hat{\cal H}_{n_A2} & \cdots & \hat{\cal H}_{n_An_A}-E 
	\end{array}
\right)
\left(
        \begin{array}{c}
        f_{1}  \\
        f_{2}  \\
         \vdots        \\
        f_{n_A}
        \end{array}
\right)
=0,
\label{eq7}
\ee
and the three-body wave function is:
\be
\Psi(\bm{x},\bm{y})=\frac{1}{\rho^{5/2}}
\left(
f_{1}, f_{2}, \cdots, f_{n_A}
\right)
\left(
        \begin{array}{c}
        \Phi_1  \\
        \Phi_2  \\
         \vdots        \\
        \Phi_{n_A}
        \end{array}
\right)
\ee

It is important to note that the diagonal terms $\hat{\cal H}_{nn}$ in Eq.(\ref{eq5}) contain the
angular eigenvalues $\lambda_n(\rho)$ introduced in Eq.(\ref{eq2}). They appear in the effective 
adiabatic potentials, which are given by:
\be
V^{(n)}_{eff}(\rho)=\frac{\hbar^2}{2m}\left( \frac{\lambda_n(\rho)+\frac{15}{4}}{\rho^2}-Q_{nn}(\rho) \right)
\label{eq9}
\ee

\begin{figure}
\vspace*{0.2cm}
\epsfig{file=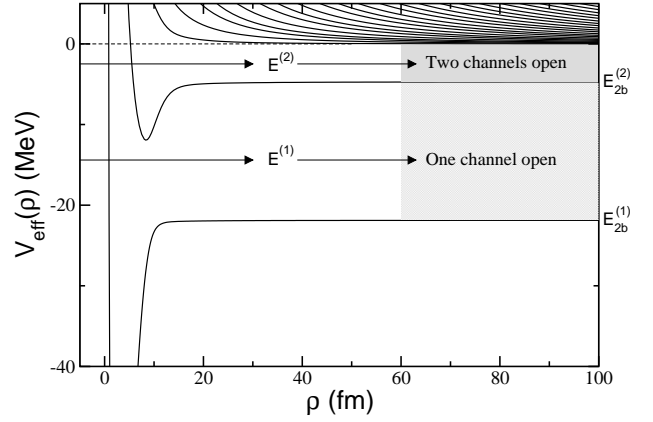, width=5.5cm, angle=-90}
\caption{ Typical effective adiabatic potentials for a three-body system where two two-body bound
states are present. The two lowest adiabatic
potentials go asymptotically to the binding energies $E_{2b}^{(1)}$ and $E_{2b}^{(2)}$ of the
two-body bound states. For a given three-body energy $E$, when $E_{2b}^{(1)}<E<E_{2b}^{(2)}$ only
one channel is open, while when $E_{2b}^{(2)}<E<0$ both channels are open.}
\label{fig1}
\end{figure}

A typical behavior of the adiabatic potentials is shown in Fig.\ref{fig1}. They correspond
to a three-body system where two of the two-body subsystems have a bound state. This is reflected
in the fact that the two lowest effective adiabatic potentials go asymptotically to the binding
energies $E_{2b}^{(1)}$ and $E_{2b}^{(2)}$ of each bound two-body system. The angular eigenfunctions
associated to these two adiabatic potentials have the general asymptotic form \cite{nie01}:
\be
\Phi_n^{JM}(\rho,\Omega)\stackrel{\rho\rightarrow \infty}{\rightarrow}\rho^{3/2} \left[ \psi_n^{j_x}(\bm{x}) 
        \otimes \left[Y_{\ell_y}(\Omega_y) \otimes \chi_{s_y} \right]^{j_y}\right]^{JM},
\label{eq10}
\ee
where for this particular case $n=1,2$ and
we have now made explicit the quantum numbers. The wave function $\psi_n^{j_x}(\bm{x})$, normalized
to 1 in the $\bm{x}$-Jacobi coordinate, describes the bound two-body system associated to the
effective potential $V_{eff}^{(n)}$, whose angular momentum is $j_x$. Asymptotically it tends to the
bound state wave function of the corresponding two-body subsystem. The spin function $\chi_{s_y}$ 
describes the spin of the third particle, which couples to the orbital angular momentum $\ell_y$
(associated to the Jacobi coordinate $\bm{y}$) to give total angular momentum $j_y$. Finally, $j_x$ and
$j_y$ couple to the total angular momentum $J$ with projection $M$ of the three-body system.

The analytic form given in Eq.(\ref{eq10}) for the asymptotic expression of the angular eigenfunction 
$\Phi_n^{JM}(\Omega,\rho)$ makes evident that it describes an
asymptotic spatial distribution for the three particles corresponding to two of them forming 
a bound state, described by $\psi_n^{j_x}(\bm{x})$, 
and a free third particle moving in the continuum. In other words, the effective
adiabatic potentials associated to angular eigenfunctions with the asymptotic form of 
Eq.(\ref{eq10}) are the ones describing
the possible incoming and outgoing channels of a process where a particle hits a bound
state formed by the other two.

In Fig.\ref{fig1} the different regions defined by the energy of the incident particles are
depicted. All the three-body energies $E$ such that 
$E_{2b}^{(1)} < E < E_{2b}^{(2)}$ (like $E^{(1)}$ in the figure) correspond to processes where 
only one channel is open. Only the elastic collision between the third particle 
and the bound two-body state with energy $E_{2b}^{(1)}$ is possible. When the three-body energy
increases up to the region $E_{2b}^{(2)} < E < 0$ ($E^{(2)}$ in the figure) a second channel is 
open. Two different collisions are now possible, the one where a particle hits the bound state 
with binding energy $E_{2b}^{(1)}$, and the one where a particle hits the state with binding 
energy $E_{2b}^{(2)}$. 
In the same way, each of these reactions has two possible outgoing channels, corresponding to 
the two allowed bound two-body states and the third particle in the continuum. In particular,
in this energy range the rearrangement process is open.
When $E>0$ the breakup channels are also open. They are described by the remaining infinitely many 
adiabatic potentials. Processes with breakup channels open will be investigated in a 
forthcoming work.

Therefore, for processes where $n_0$ channels are open, the full three-body wave function 
has actually $n_0$ different components. We shall denote them by $\Psi_i$, corresponding to the process 
with incident channel $i$. Each of the three-body functions $\Psi_i$ is then expanded
as in Eq.(\ref{eq3}), but the radial functions need now an additional index $i$ ($i \leq n_0$) indicating 
the incident channel to which they correspond: 
\be
\Psi_i = \frac{1}{\rho^{5/2}}\sum_{n=1}^{n_A} 
f_{ni}(\rho) \Phi_n(\rho,\Omega).
\label{eq11}
\ee

For each open channel $i$ ($i=1,2,\cdots,n_0$) the corresponding radial wave functions 
$f_{ni}(\rho)$ satisfy the 
set of radial equations of Eq.(\ref{eq4}), and Eq.(\ref{eq7}) can be generalized to
\begin{small}
\be
\left(
        \begin{array}{ccc}
        \hat{\cal H}_{11}-E &  \cdots & \hat{\cal H}_{1n_A} \\
        \hat{\cal H}_{21} &  \cdots & \hat{\cal H}_{2n_A} \\
         \vdots       &  \vdots & \vdots        \\
        \hat{\cal H}_{n_A1} & \cdots & \hat{\cal H}_{n_An_A}-E
        \end{array}
\right)
\left(
        \begin{array}{cccc}
        f_{11}  & f_{12}  & \cdots   & f_{1n_0}\\
        f_{21}  & f_{22}  & \cdots   & f_{2n_0}\\
         \vdots & \vdots  & \vdots   & \vdots \\
        f_{n_A1}& f_{n_A2}& \cdots   & f_{n_An_0}
        \end{array}
\right)
=0,
\label{eq12}
\ee
\end{small}
which describes the full process. The full three-body wave function
is now given by:
\begin{small}
\be
\Psi=
\left(
        \begin{array}{c}
        \Psi_1  \\
        \Psi_2  \\
         \vdots        \\
        \Psi_{n_0}
        \end{array}
\right)
= \frac{1}{\rho^{5/2}}
\left(
        \begin{array}{cccc}
        f_{11}  & f_{21}  & \cdots   & f_{n_A1}\\
        f_{12}  & f_{22}  & \cdots   & f_{n_A2}\\
         \vdots & \vdots  & \vdots   & \vdots \\
        f_{1n_0}& f_{2n_0}& \cdots   & f_{n_An_0}
        \end{array}
\right)
\left(
        \begin{array}{c}
        \Phi_1  \\
        \Phi_2  \\
         \vdots        \\
        \Phi_{n_A}
        \end{array}
\right)
\label{eq13}
\ee
\end{small}

\subsection{Asymptotics: ${\cal K}$-matrix and ${\cal S}$-matrix}

For scattering states and energies below the breakup threshold ($E<0$),
the Eqs.(\ref{eq12}) decouple asymptotically, and for a given incident
channel $i$ ($i=1,\cdots,n_0$) the only equations surviving
are the ones of the form:
\be
(\hat{\cal H}_{nn}-E)f_{ni}(\rho)=0 \hspace*{1cm}(n=1,\cdots,n_A),
\ee
which, by use of Eq.(\ref{eq5}), can be written as:
\be
\left(-\frac{\hbar^2}{2 m} \frac{d^2}{d\rho^2} +V_{eff}^{(n)}(\rho)-E \right)
 f_{ni}(\rho)=0,
\ee
where $V_{eff}^{(n)}$ is given by Eq.(\ref{eq9}).

When $n$ corresponds to a closed channel, the radial wave functions $f_{ni}$ vanish 
asymptotically. When $n$ corresponds to an open channel, the asymptotic behavior 
of $f_{ni}$ is dictated by the asymptotics of the corresponding adiabatic 
potential $V_{eff}^{(n)}$.
A careful analysis of the large distance behavior of the $\lambda_n(\rho)$ and 
$Q_{nn}(\rho)$ functions in the case of bound two-body subsystems can be found 
in Ref.~\cite{nie01}. In particular, Eqs.(91) and (93) of that reference allow to rewrite 
the above equation for the case $n\leq n_0$ as:
\be
\left[\frac{d^2}{d\rho^2}+(k_y^{(n)})^2-\frac{\ell_y(\ell_y+1)}{\rho^2}
\right] f_{ni}(\rho)=0
\label{eq16}
\ee
where
\be
k_y^{(n)}=\sqrt{ \frac{2m}{\hbar^2}(E-E_{2b}^{(n)}) },
\ee
$E_{2b}^{(n)}$ is the binding energy of the bound two-body system associated to the
open channel $n$, and $\ell_y$ is the orbital angular momentum associated to the
Jacobi coordinate $\bm{y}$, which amounts to the relative orbital angular momentum
between the projectile and the two-body bound target.

 From Eq.(\ref{eq16}) it is now clear that the asymptotic behavior of the
$f_{ni}$ functions ($n,i\leq n_0$) is given by:
\be
f_{ni}(\rho) \rightarrow \sqrt{k_y^{(n)}}\rho
\left( A_{in}^{(K)} j_{\ell_y}(k_y^{(n)}\rho) + B_{in}^{(K)} \eta_{\ell_y}(k_y^{(n)}\rho) \right)
\label{eq18}
\ee
where $j_{\ell_y}$ and $\eta_{\ell_y}$ are the usual regular and irregular spherical Bessel functions,
respectively. The superscript $(K)$ indicates that, with this particular choice, the 
coefficients $A_{in}^{(K)}$ and $B_{in}^{(K)}$ will permit to extract the ${\cal K}$-matrix. Conversely,
using the spherical Hankel functions in Eq.(\ref{eq18}) the coefficients will form
the ${\cal S}$-matrix and the superscript $(S)$ will be used (see below).

Therefore, asymptotically, the matrix containing the radial wave functions in Eq.(\ref{eq13}) 
reduces to the $n_0\times n_0$ matrix $A^{(K)} J + B^{(K)} Y$, 
where $A^{(K)}$ and $B^{(K)}$ are $n_0\times n_0$ matrices whose components are the 
$A_{ij}^{(K)}$ and $B_{ij}^{(K)}$ coefficients
of Eq.(\ref{eq18}), and $J$ and $Y$ are two $n_0\times n_0$ diagonal
matrices with diagonal terms 
$\left( \sqrt{k_y^{(i)}}\rho j_{\ell_y}(k_y^{(i)}\rho) \right)$ and
$\left( \sqrt{k_y^{(i)}}\rho \eta_{\ell_y}(k_y^{(i)}\rho) \right)$, respectively.
Thus, the asymptotic behavior of the full three-body wave function (\ref{eq13}) 
can be finally written as:
\be
\Psi \rightarrow A^{(K)} F_\rho^{(K)} + B^{(K)} G_\rho^{(K)},
\label{eq20}
\ee
where $F_\rho^{(K)}$ and $G_\rho^{(K)}$ are column vectors with $n_0$ terms of the form 
$\left( \sqrt{k_y^{(n)}} j_{\ell_y}(k_y^{(n)}\rho) \Phi_n /\rho^{3/2} \right)$
and 
$\left( \sqrt{k_y^{(n)}} \eta_{\ell_y}(k_y^{(n)}\rho) \Phi_n /\rho^{3/2} \right)$,
respectively.

 From Eq.(\ref{eq20}) we then have that for a given incident channel $i$ the 
asymptotic form of the corresponding three-body wave function (\ref{eq11}) takes the form:
\be
\Psi_i \rightarrow
\sum_{n=1}^{n_0}
\left(
A_{in}^{(K)} F_{\rho,n}^{(K)} + B_{in}^{(K)} G_{\rho,n}^{(K)}
\right)
\label{eq21}
\ee
where
\begin{eqnarray}
F_{\rho,n}^{(K)}&\!\!\!\!=\!\!\!\!&\sqrt{k_y^{(n)}}j_{\ell_y}(k_y^{(n)}\rho)
\left[ \psi_n^{j_x} \otimes
\left[Y_{\ell_y}(\Omega_y) \otimes \chi_{s_y} \right]^{j_y}\right]^{JM}
\nonumber
\\
G_{\rho,n}^{(K)}&\!\!\!\!=\!\!\!\!&\sqrt{k_y^{(n)}}\eta_{\ell_y}(k_y^{(n)}\rho)
\left[ \psi_n^{j_x} \otimes
\left[Y_{\ell_y}(\Omega_y) \otimes \chi_{s_y} \right]^{j_y}\right]^{JM}
\label{eq21b}
\end{eqnarray}
and where we have made use of Eq.(\ref{eq10}), which relates the angular eigenfunction $\Phi_n^{JM}$
and the two-body wave function $\psi_n^{j_x}$. When two or three 
identical particles are present in the system, these functions should be correctly 
symmetrized or antisymmetrized depending on whether they are either bosons or fermions. 

 From Eq.(\ref{eq20}) we can now easily write:
\be
\Psi \rightarrow A^{(K)} \left(F_\rho^{(K)} - {\cal K} G_\rho^{(K)}\right),
\label{eq23}
\ee
where
\be
{\cal K}= -{A^{(K)}}^{-1} B^{(K)}
\label{eq23b}
\ee
is the ${\cal K}$-matrix of the reaction, whose dimension is $n_0 \times n_0$  (with $n_0$ being the number
of open channels).

The discussion in this subsection could have also been made by replacing $j_{\ell_y}$ and 
$\eta_{\ell_y}$  in Eq.(\ref{eq18}) by the spherical Hankel functions $h_{\ell_y}^{(2)}$  and
$h_{\ell_y}^{(1)}$, respectively. This would then lead to:
\be
\Psi \rightarrow A^{(S)} F_\rho^{(S)} + B^{(S)} G_\rho^{(S)},
\label{eq24}
\ee
where now $F_\rho^{(S)}$ and $G_\rho^{(S)}$ are column vectors with $n_0$ terms of the form
$\left( \sqrt{k_y^{(n)}} h_{\ell_y}^{(2)}(k_y^{(n)}\rho) \Phi_n /\rho^{3/2} \right)$
and
$\left( \sqrt{k_y^{(n)}} h_{\ell_y}^{(1)}(k_y^{(n)}\rho) \Phi_n /\rho^{3/2} \right)$,
respectively. We can then write:
\be
\Psi \rightarrow A^{(S)} \left(F_\rho^{(S)} + {\cal S} G_\rho^{(S)}\right),
\ee
where
\be
{\cal S}={A^{(S)}}^{-1}B^{(S)}
\label{eq26}
\ee
is the so called ${\cal S}$-matrix of the reaction. 
The ${\cal S}$ and ${\cal K}$ 
matrices are related through the well known simple expression:
\be
{\cal S}=(1+i{\cal K})(1-i{\cal K})^{-1}.
\label{eq27}
\ee

It is important to keep in mind that while $A^{(K)}$, $B^{(K)}$, and ${\cal K}$ are real, the
matrices $A^{(S)}$, $B^{(S)}$, and ${\cal S}$ are in general complex.

\section{Second order integral relations}

In Ref.\cite{bar09} the applicability of the HA expansion to extract phase 
shifts for 1+2 reactions when only the elastic channel is open has been
discussed. In that reference it was found that, when increasing the number of adiabatic 
channels $n_A$ included in the calculation as much as possible, the 
difference between the computed phase shift and the exact value remains significant. 
As mentioned in the Introduction, this is related
to the fact that the asymptotic structure of the system has to be describe in
terms of spherical Bessel functions depending on $k_y y$, where $y$ is 
the modulus of the Jacobi
coordinate between the center of mass of the outgoing bound two-body system and the
third particle.  Instead, the asymptotic behavior using the HA basis is given in terms
of spherical Bessel functions depending on $k_y \rho$.
Since the equivalence between $k_y y$ and $k_y \rho$
is not matched for any finite value of $\rho$, the correct boundary condition
is only achieved at $\rho \approx \infty$ and $n_A \rightarrow \infty$.

For a general multichannel process the adiabatic expansion obviously shows
the same deficiency. The correct asymptotic wave function is given by Eq.(\ref{eq20}),
but where $F_\rho^{(K)}$ and $G_\rho^{(K)}$ in Eq.(\ref{eq21b}) have to be replaced by $F^{(K)}$ and $G^{(K)}$,
which are column vectors whose $n$-th element is
\begin{eqnarray}
F_n^{(K)}&\!\!\!\!=\!\!\!\!&\sqrt{k_y^{(n)}}j_{\ell_y}(k_y^{(n)}y_n)
\left[ \psi_n^{j_x} \otimes
\left[Y_{\ell_y}(\Omega_y) \otimes \chi_{s_y} \right]^{j_y}\right]^{JM}
\nonumber
\\
G_n^{(K)}&\!\!\!\!=\!\!\!\!&\sqrt{k_y^{(n)}}\eta_{\ell_y}(k_y^{(n)}y_n)
\left[ \psi_n^{j_x} \otimes
\left[Y_{\ell_y}(\Omega_y) \otimes \chi_{s_y} \right]^{j_y}\right]^{JM}.
\label{eq28}
\end{eqnarray}
In these expressions $y_n$ refers to the modulus of the Jacobi coordinate describing the center
of mass of the bound two-body system $\psi_n^{j_x}$ and the third particle.

It is important to recall that the Bessel functions $\eta_\ell$ are irregular at the origin, which
creates difficulties from the numerical point of view. It is then convenient to regularize such
function, in such a way that $G_n^{(K)}$ given in Eq.(\ref{eq28}) has to be replaced by:
\be
\widetilde{G}_n^{(K)}=\left(1-e^{-\gamma y_n} \right)^{\ell_y+1}G_n^{(K)}
\label{eq28b}
\ee
where $\gamma$ is a non linear parameter. The results are stable for values of $\gamma$ within a small 
range around  $\gamma\sim 1/r_0$, with $r_0$ the range of the potential.

 For simplicity in the notation, from now on, we shall refer to the matrices 
$\{F^{(K)}, \widetilde{G}^{(K)}\}$  as $\{F, G\}$, in such a way that we can write the asymptotic 
behavior of the wave function as:
\be
\Psi \rightarrow A F + B G,
\label{eq29}
\ee
and ${\cal K}=-A^{-1} B$.  

The vectors $F$ and $G$ satisfy the following normalization condition:
\be
-\frac{2m}{\hbar^2}
\left[
  \langle F|\hat{\cal H}-E |G \rangle - \langle G | \hat{\cal H}-E |F \rangle^T
\right]=\mathbbm{I}.
\label{eq37}
\ee
where $\mathbbm{I}$ is the identity matrix.  
In Eq.(\ref{eq37}) we have introduced a notation to be used from now on
in which the overlap of two vectors is a matrix whose elements are, for example,
$(\langle F|\hat{\cal H}-E |G \rangle)_{ij}=\langle F_i|\hat{\cal H}-E |G_j \rangle$.
The normalization condition allows to extract a first order estimate of
the matrices $A$ and $B$ from the scattering wave function $\Psi$ as
\begin{eqnarray}
        B^{1^{st}} & = & -\frac{2m}{\hbar^2} 
\left[
  \langle F|\hat{\cal H}-E |\Psi \rangle^T - \langle\Psi | \hat{\cal H}-E |F 
\rangle \right]  \label{eq35}\\
        A^{1^{st}} & = & -\frac{2m}{\hbar^2} 
\left[
  \langle \Psi|\hat{\cal H}-E |G \rangle - \langle G | \hat{\cal H}-E |\Psi 
\rangle^T
\right]. 
\label{eq36}
\end{eqnarray}
Clearly, when $\Psi$ is an exact solution of $(\hat{\cal H}-E)\Psi=0$, the above
expressions reduce to the the following integral relations:
\begin{eqnarray}
	B & = &   \frac{2m}{\hbar^2} 
   \langle\Psi | \hat{\cal H}-E |F \rangle  \nonumber \\
	A & = & -\frac{2m}{\hbar^2} 
  \langle \Psi|\hat{\cal H}-E |G \rangle\label{eq38}. 
\end{eqnarray}
Explicitly, each matrix element $B_{ij}$ and $A_{ij}$ is given by:
\begin{eqnarray}
	B_{ij} & = &  \frac{2m}{\hbar^2} 
   \langle\Psi_i | \hat{\cal H}-E |F_j \rangle \label{eq39} \\
	A_{ij} & = & - \frac{2m}{\hbar^2} 
  \langle \Psi_i|\hat{\cal H}-E |G_j \rangle \label{eq40}, 
\end{eqnarray}
which can be seen as the extension to multichannel scattering of the expressions valid for the
single channel case. Now the same formula applies for each possible 
incoming channel described by $\Psi_i$ and each possible outgoing channel whose asymptotic
analytic form is given by a linear combination of $F_j$ and $G_j$.	

As demonstrated in Refs.~\cite{bar09,kiev10} for a single channel process, the relation 
${\cal K}=-A^{-1} B$  computed using Eqs.(\ref{eq39}) and (\ref{eq40})
can be considered accurate up to second order when a trial 
wave function $\Psi_t$ is used. Moreover the two integral relations 
of Eqs.(\ref{eq39}) and (\ref{eq40}) can be directly derived from the 
Kohn Variational Principle. As shown in Appendix A, the matrix form of
of KVP, necessary to describe a multichannel process,  establishes that each matrix element
of $A^{-1}B^{2^{nd}}$ is a functional given by
\be
A^{-1}B^{2^{nd}}=A^{-1}B+\frac{2m}{\hbar^2} 
A^{-1} \langle \Psi_t | \hat{\cal H}-E | \Psi_t \rangle(A^{-1})^T \;,
\label{eq41}
\ee
which is stationary with respect to variations of the wave function.
Taking into account the general asymptotic behavior in Eq.(\ref{eq29}), 
we can write the full trial wave function schematically as:
\be
\Psi_t=\Psi_c+AF+BG,
\label{eq42}
\ee
with $\Psi_c \rightarrow 0$ as $\rho \rightarrow \infty$. Furthermore
$\Psi_c$ can be expanded in 
terms of a (square integrable) complete basis $\{b_i, i=1,\cdots,m \}$:
\begin{small}
\begin{equation}
\Psi_c=
\left(
        \begin{array}{c}
        \Psi_{c,1}  \\
        \Psi_{c,2}  \\
         \vdots        \\
        \Psi_{c,n_0}
        \end{array}
\right)
= \frac{1}{\rho^{5/2}}
\left(
        \begin{array}{cccc}
        c_{11}  & c_{12}  & \cdots   & c_{1m}\\
        c_{21}  & c_{22}  & \cdots   & c_{2m}\\
         \vdots & \vdots  & \vdots   & \vdots \\
        c_{n_01}& c_{n_02}& \cdots   & c_{n_0m}
        \end{array}
\right)
\left(
        \begin{array}{c}
        b_1  \\
        b_2  \\
         \vdots        \\
        b_m
        \end{array}
\right),
\end{equation}
\end{small}
The variation of the functional with respect to the linear parameters
$c_{ij}$ and with respect to the matrix elements of $A^{-1}B$ leads to:
\begin{eqnarray}
\langle \Psi_c | \hat{\cal H}-E | \Psi_t \rangle &= & 0
\nonumber \\
\langle G | \hat{\cal H}-E | \Psi_t \rangle &= & 0
\label{eq44}
\end{eqnarray}
When $\Psi$ is replaced by $\Psi_t$, the second expression above and Eq.(\ref{eq36}) result:
\begin{equation}
        A  =  -\frac{2m}{\hbar^2} 
  \langle \Psi_t|\hat{\cal H}-E |G \rangle. 
\end{equation}
Replacing now Eq.(\ref{eq42}) into (\ref{eq41}), and making use of the Eqs.(\ref{eq44}),
 we also get:
\be
B^{2^{nd}}=B^{1^{st}}+ \frac{2m}{\hbar^2}
\langle F | \hat{\cal H}-E | \Psi_t \rangle^T,
\ee
and, taking into account that $B^{1^{st}}$ is given by Eq.(\ref{eq35}) 
we can then obtain the final result:
\begin{eqnarray}
        B^{2^{nd}} & = &  \frac{2m}{\hbar^2} 
   \langle\Psi_t | \hat{\cal H}-E |F \rangle  \nonumber \\
        A & = & -\frac{2m}{\hbar^2} \langle \Psi_t|\hat{\cal H}-E |G \rangle, 
\label{eq47}
\end{eqnarray}
which according to Eqs.(\ref{eq23b}) and (\ref{eq27}) permit to obtain 
the second order estimate of the ${\cal K}$-matrix or the ${\cal S}$-matrix,
${\cal K}^{2^{nd}}$ or ${\cal S}^{2^{nd}}$, respectively.

In practical cases, application of the integral relations given in Eq.(\ref{eq47}) 
require the calculation of each individual matrix element $A_{ij}$ and $B_{ij}$, 
which relate each possible incoming channel $i$ described by $\Psi_i$, with each 
possible outgoing asymptotics given by $F_j$ and $G_j$. Details about the calculation of these 
matrix elements are given in appendix B, in particular for the case of two-body potentials 
projecting on the partial waves. 

The integral relations of Eq.(\ref{eq47}) depend on the
short range structure of the scattering wave function $\Psi_t$ as $F$ and $G$ are
asymptotically solutions of $({\cal H}-E)F,G=0$. This property allows for different
applications of the integral relations, as discussed in Ref.~\cite{kiev10}. In the
present work the interest is given in the study of the pattern of convergence of
${\cal K}^{2^{nd}}$ in terms of the number of equations $n_A$ considered in the description
of $\Psi_t$ using the HA expansion. As we will see, increasing $n_A$, both matrices
$A$ and $B^{2^{nd}}$ slightly change, showing individually a very slow rate of convergence. 
Conversely, its rate ${\cal K}^{2^{nd}}=-A^{-1}B^{2^{nd}}$ shows a pattern of 
convergence similar to that one observed in a bound state calculation.

\section{Results}

\subsection{Test of the method: A model 1+2 collision}
\label{sec2a}

\begin{figure}
\vspace*{0.2cm}
\epsfig{file=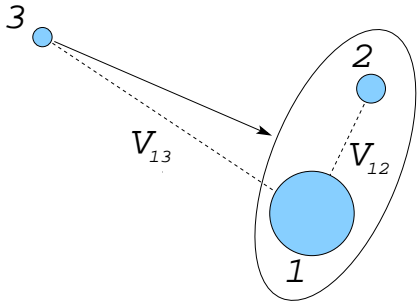, width=5.5cm, angle=0}
\caption{ Scheme of the model reaction used to test the integral relations. The light projectile
3 hits the dimer made by particles 1 and 2. Particle 1 is infinitely heavy and particles 2 and 3
do not interact. }
\label{fig1b}
\end{figure}

To test the method we have chosen a 1+2 reaction where the target dimer is made by an infinitely 
heavy particle and a light one, and where we consider a projectile interacting only with the heavy 
particle (see Fig.\ref{fig1b}). In the collision particle 2 does not play any role, and the process is 
equivalent to a two-body reaction between particles 3 and 1. Therefore the results obtained
through the three-body calculation and the integral relations can be easily tested by means of
a simple two-body calculation.

In particular, we consider a two-body target made by two spin-zero bosons 
with masses $0.5 m$ and $10^{12} m$ (with $m=938.69461$ MeV) interacting via a simple central 
potential given by

\begin{equation}
V_{12}(r)=-80 \;e^{-r^2/1.6^3}
\end{equation}
where $r$ is given in fm and the strength in MeV. This system has only one $s$-wave
bound state with binding energy $-6.2757$ MeV. 

The projectile, which is chosen to have a mass of  $0.51 m$, does not interact with particle 2, while
it does it with particle 1 through the gaussian potential:
\begin{equation}
V_{13}(r)=-30\; e^{-r^2/1.6^2},
\end{equation}
where again $r$ is in fm and the strength in MeV. This potential is not able to bind particles 1 and 3.
Finally, as described above, $V_{23}=0$.

We have chosen an incident energy of $3$ MeV, which implies a total three-body 
energy of $-3.2757$ MeV. We are then below the threshold for breakup of the two-body
target, and only the elastic channel is open. Therefore, $B$ and $A$ in Eq.(\ref{eq47}) are
just numbers, and they are such that $\tan \delta_\ell=-B/A$ (note that the definition of $B$
in here and in \cite{bar09,kiev10} have opposite sign). 

\begin{table}
\caption{Partial wave phase shifts $\delta_{\ell}$ for different values 
of $n_A$ (number of adiabatic terms used in the expansion (\ref{eq3})).  
In the last row, the result using a two-body calculation is shown.
}
\begin{ruledtabular}
\begin{tabular}{c c c c}
 $n_A$         & $\delta_{s}$ & $\delta_{p}$ & $\delta_{d}$ \\
\hline
1              & 40.554       &  0.6658      & 0.0136\\
2              & 38.988       &  0.6892      & 0.0113\\
3              & 38.642       &  0.6921      & 0.0121\\
5              & 38.693       &  0.6911      & 0.0119\\
8              & 38.702       &  0.6918      & 0.0118\\
10             & 38.701       &  0.6918      & 0.0118\\
\hline
two-body       &\bm{38.699}   & \bm{0.6917}   & \bm{0.0117} \\
\end{tabular}
\end{ruledtabular}
\label{tab1}
\end{table}

We have computed the phase shift for this reaction for relative $s$, $p$, and $d$ waves between
the projectile and the target. The convergence of the expansion (\ref{eq3})  
is shown in table \ref{tab1}, where we show the phase shift for the different partial waves and 
for different values of $n_A$, which is the number of adiabatic terms included in the
calculation. As we can see, inclusion of 8 to 10 adiabatic potentials is enough to reach
convergence for the three partial waves. Furthermore, the converged result agrees with the phase shift
obtained from the two-body calculation describing the collision between particles 3 and 1.

\begin{figure}
\vspace*{0.2cm}
\epsfig{file=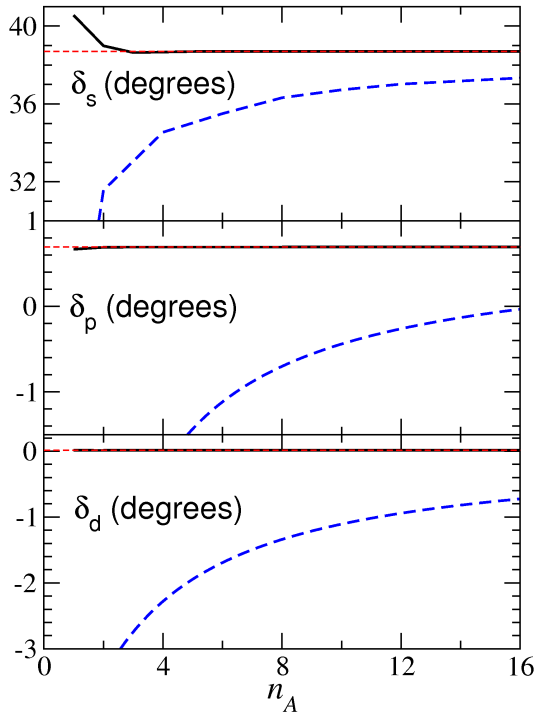, width=7.5cm, angle=0}
\caption{(color online) Phase-shift for $s$, $p$, and $d$ partial waves for the model reaction 
in section ~\ref{sec2a} as a function of the number of adiabatic terms included in the calculation. 
The solid line is the result obtained through the integral relations, and the thick dashed curve has 
been obtained from the asymptotic expression (\ref{eq18}). The thin dashed line is the result obtained from a 
two-body calculation.} 
\label{fig-del}
\end{figure}

The efficiency of using the integral relations in Eq.(\ref{eq47}) is made evident in Fig.~\ref{fig-del},
where we show the partial wave phase shifts $\delta_\ell$ as a function of $n_A$. The solid line
gives the results obtained from the integral relations (given in table~\ref{tab1}), and the thick
dashed line shows the results extracted by direct comparison of the computed asymptotic radial wave
functions and the analytic expression in Eq.~(\ref{eq18}). The thin dashed line indicates the phase shifts
obtained from a two-body calculation. As we can immediately see in the figure, the pattern of convergence 
of the phase shifts obtained from Eq.(\ref{eq18}) (thick dashed curves) is very slow. A simple extrapolation 
of these curves up to the correct value permits to foresee that the number of adiabatic terms needed to 
obtain accurate values of $\delta_\ell$ is far larger than the one needed when the integral relations are used.
In fact, at the scale of the figure, the calculations with the integral relations are already for $n_A=4$
indistinguishable from the correct result (see also table~\ref{tab1}).

\begin{figure}
\vspace*{0.2cm}
\epsfig{file=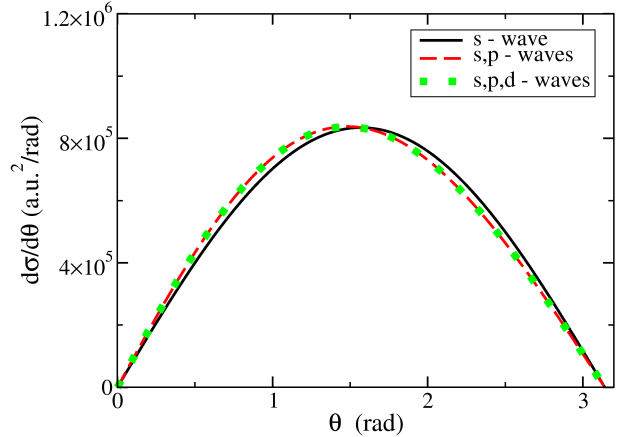, width=8.5cm, angle=0}
\caption{(color online) Cumulative contributions to the
differential cross section as a function of the scattering angle $\theta$ for
the model collision in section~\ref{sec2a}.} 
\label{fig2}
\end{figure}

As seen in table~\ref{tab1}, the $d$-wave phase shift is already rather small and therefore, at the 
considered energy, the cross section contributions from higher angular partial waves is negligible. 
In Fig. \ref{fig2} 
the differential cross section of the process with cumulative inclusion of
one (solid), two (dashed), and three (dot-dashed)
partial waves is shown. As can be seen, $s$-, $p$ and $d$-partial waves are
enough to obtain a converged cross section of the process. In fact, the partial waves beyond the 
$s$-wave have a modest contribution, and the total cross section approaches quite a lot the 
characteristic $\sin \theta$ function of the $s$-waves.

\subsection{A realistic case: The \he-\dimer collision}

In this section we discuss an interesting simple physical case, but technically similar to 
the model case described in the previous section. This is the collision of a \he atom into 
the weakly bound \dimer  dimer.
The helium dimer has a single $s$-wave bound state, and as soon as the incident energy is below the dimer
breakup threshold, again only the elastic channel is open.

The two-body helium-helium interaction is chosen to be the simple effective gaussian potential 
given in \cite{nil98}. This is enough for our purpose of illustrating how this kind of processes can be
easily described by use of the integral relations. This potential is built to reproduce the $s$-wave 
scattering length (189.054 a.u.) and effective range (13.843 a.u.) of the LM2M2 interaction \cite{azi91}, 
and it is given by:
\begin{equation}
V_{2B}(r)=-1.227 \; e^{-r^2/10.03^2}
\end{equation}
where $r$ is given in a.u. and the strength is in K. This potential leads to a bound 0$^+$
\dimer dimer with a binding energy  $E_d=-1.2959$ mK, a scattering length $a=189.947$ a.u.,
and an effective range of 13.846 a.u.. Simple representations of the atom-atom
potentials are often used to describe reactions in the ultracold regime (see for example
Refs.~\cite{incao08,stecher09}). In this regime the process is largely independent of
the shape of the potential and can be characterized only by the scattering length.

With this interaction the helium trimer has two bound states at $-150.0$ mK and $-2.467$ mK.
These states have been obtained using the gaussian potential active only in $s$-waves.
Increasing the number of partial waves up to $\ell_x=\ell_y=8$ results in a very small change 
for the ground and excited state binding energies, which become now $150.4$ mK and $-2.472$ mK,
respectively. In fact, more than 99\% of the norm of the bound state wave functions is provided 
by the lowest adiabatic
term, whose corresponding adiabatic potential is close to identical in both calculations. 
Accordingly, in the following we restrict the calculations to include only the $\ell_x=0$ channel.

When the LM2M2 potential is used, these two states are found to have binding energies 
$-126.4$ mK and $-2.265$ mK, respectively \cite{bar01}. As we can
see, the ground state is not very well reproduced when the gaussian version
of the potential is used. In this very deep state, the three atoms are close to
each other and the correct structure can not be described with the simplified
potential. Conversely, the excited state which has the characteristic of an
Efimov state has an structure in which the third atom orbits very far from the bound
state of the other two. This particular structure is well described by the
attractive gaussian potential.

\begin{table}
\caption{Partial wave phase shifts $\delta_{\ell}$ are given
showing convergence of these values
with the number $n_A$ of adiabatic potential used in the calculation.
The last row shows the result when the Hyperspherical Harmonic method \cite{kiev08} is used.
}
\begin{ruledtabular}
\begin{tabular}{c c c c c}
 $n_A$         & $\delta_{s}$ & $\delta_{p}$ & $\delta_{d}$ & $\delta_{f}$ \\
\hline
1             & -39.72  &  -13.19  & 2.01  & -0.27\\
2             & -40.30  &  -13.13  & 2.11  & -0.28\\
4             & -40.43  &  -13.11  & 2.13  & -0.28\\
8             & -40.50  &  -13.11  & 2.14  & -0.28\\
18            & -40.54  &  -13.11  & 2.14  & -0.28\\
22            & -40.54  &  -13.11  & 2.14  & -0.28\\
\hline
HH-calculation       &\bm{-40.55}   & ---   & --- & --- \\
\end{tabular}
\end{ruledtabular}
\label{tab2}
\end{table}

In order to study the convergence properties of the HA expansion for a $1+2$ collision,
we have chosen an incident energy of $0.5$ mK (or a three-body energy $E=-0.7959$ mK). The 
phase shifts for the different partial waves have been computed as in the previous subsection. 
The results are shown in table \ref{tab2} for $s$, $p$, $d$, and $f$ waves. A good convergence 
is obtained already after inclusion of about 10 adiabatic terms, except for $s$-waves, where about
18 are needed. The last row in the table shows the phase shift obtained for an $s$-wave collision when
the Hyperspherical Harmonic method is used. The two methods are in close agreement.

\begin{figure}
\vspace*{0.2cm}
\epsfig{file=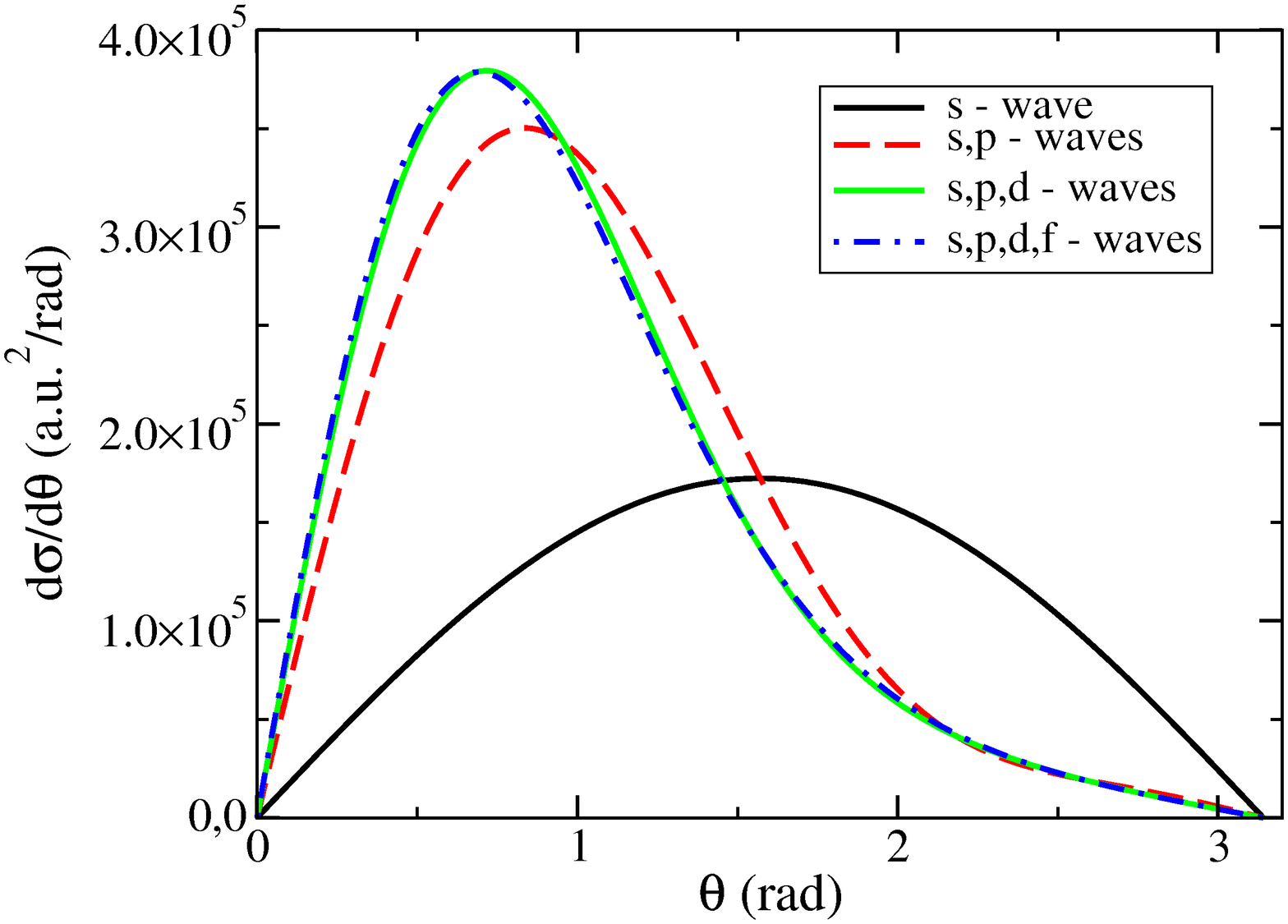, width=8.5cm, angle=0}
\caption{(color online) The same as in Fig.\ref{fig2} for the \he-\dimer collision. With four partial waves 
the convergence is fairly good.  }
\label{fig3}
\end{figure}

At this particular energy we have calculated the differential cross section.
Fig. \ref{fig3} shows the cumulative contributions of the $s$, $p$, $d$ and $f$ partial
waves. We observe that the $p$-wave
contribution is rather important and produces a deviation 
from the $\sin \theta$ shape. Moreover, four partial waves are needed to reach a 
good convergence. 

\begin{figure}
\vspace*{0.2cm}
\epsfig{file=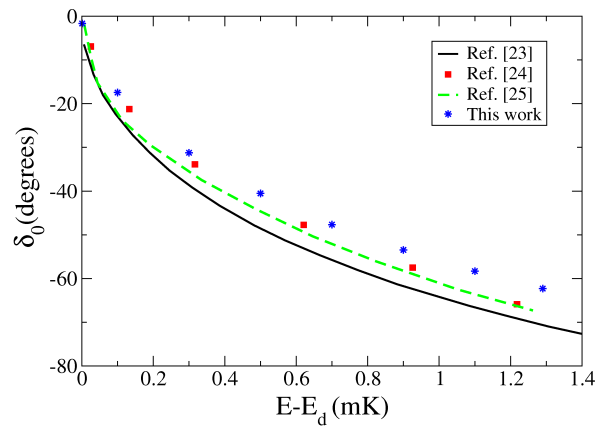, width=8cm, angle=0}
\caption{(color online)$s$-wave phase shift $\delta_s$ as a function of the incident energy
for the \he-\dimer collision. The stars are the results obtained in the present work. The
solid curve, the squares, and dashed curve correspond to the results given in \cite{sun08}, \cite{rou03}, 
and \cite{mot01}, respectively.}
\label{figcomp}
\end{figure}

In Fig.~\ref{figcomp} we show the computed $s$-wave phase shift as a function of the incident
energy ($E-E_d$). Our results are given by the stars. For comparison we also show the results
reported in \cite{sun08}, \cite{rou03}, and \cite{mot01} (solid curve, squares, and dashed curve, 
respectively). As we can see, the phase shifts 
obtained in this work are a few degrees above the ones obtained in the previous calculations, where 
the \he-\he interaction is treated more in detail. In fact the reason for this discrepancy is the 
hard core repulsion present in the \he-\he interactions used in \cite{sun08,rou03,mot01}. For the 
same reason the atom-dimer scattering length $a_{a-d}$ obtained with the gaussian potential used
in this work, $a_{a-d}=166$ a.u., differs from the typical values of around $a_{a-d}=220$ a.u. 
($\approx 116$ \AA) obtained when hard core potentials are used \cite{rou03,kol04}.  

\subsection{A multichannel collision: The $^4$He-$^4$He-$^6$Li system}

In this subsection a reaction where more than one channel is open is discussed. To this
aim, we have chosen a process involving two helium and one lithium atoms. 
The cross section for this kind of reactions is the necessary ingredient
to obtain the recombination rate for such three-body systems. As quoted in \cite{suno09}, 
where the three-body recombination for cold helium-helium-alkali-metal
systems is investigated, such collision processes are important in ultracold gas experiments
 using buffer-gas cooling, since it might limit the lifetimes of the trapped atoms.

To describe this three-body system
we take the same helium-helium interaction as in the previous section, which leads to a
0$^+$ \dimer dimer with a binding energy of $-1.2959$ mK. The lithium-helium interaction is also
chosen to have a gaussian shape, and it is taken to be:
\begin{equation}
V_{(^6{\rm Li}-^4{\rm He})}(r)=-0.27368 \; e^{-r^2/20.14^2},
\end{equation}
where $r$ is in a.u. and the strength is in K. The parameters have been adjusted to
give a scattering length of $-173.5$ a.u. and an effective range of
$26.475$ a.u. in agreement with the values obtained 
in \cite{kle99} ($a=-173.8$ a.u. and $r_e$=26.483 a.u.), where the more sophisticated 
KTTY potential is used. This potential leads to a 0$^+$ bound $^6$Li-$^4$He system with a binding energy 
of $-1.4225$ mK. 

The adiabatic potentials obtained for the $^4$He-$^4$He-$^6$Li three-body system follow the 
same pattern as the potentials in Fig.\ref{fig1}, where $E_{2b}^{(1)}$ corresponds now to the 
binding energy of the $^6$Li-$^4$He dimer ($-1.4225 $ mK) and $E_{2b}^{(2)}$ corresponds to the 
binding energy of the \dimer dimer ($-1.2959 $ mK). The three-body system presents one
bound state at $E=-58.12$ mK.

Thus, as soon as the three-body energy
lies in the same region as $E^{(2)}$ in the figure, two different channels are open. 
One of them corresponds
to a bound $^6$Li-$^4$He dimer and the second $^4$He atom in the continuum (we shall refer to it 
as channel 1), and the other one corresponds to the bound \dimer dimer and the $^6$Li atom in the 
continuum (we shall refer to it as channel 2). In other words, when taking channel 1 as the 
incoming channel we are considering a process were the $^4$He atom hits a bound $^6$Li-$^4$He dimer, 
while when choosing channel 2 as the incoming channel we are then considering the process of a 
$^6$Li atom hitting a \dimer dimer. For each of the two possible incoming channels we have two
different outgoing channels, the elastic one and a rearrangement process 
where the projectile is captured by one of the constituents of the dimer, while the
second dimer constituent is released.

The existence of two open channels implies that the ${\cal K}$-matrix
(or the ${\cal S}$-matrix) is a 2$\times$2 matrix, that can be obtained through the 
2$\times$2 matrices $A$ and $B^{2^{nd}}$ in Eq.(\ref{eq47}). Each of the four terms
in $A$ and $B^{2^{nd}}$ can be obtained as in Eqs.(\ref{eq40}) and (\ref{eq39})
where $\Psi_i$ is the trial three-body wave function for the incoming channel 
$i$. $F_j$ and $G_j$ are the asymptotic functions given in Eqs.(\ref{eq28}) 
for the outgoing channel $j$. These functions have to be symmetrized when the 
outgoing channel is 1, since the helium atom in the dimer is identical to the 
one moving in the continuum. For outgoing channel 2 this is not necessary, since the
dimer wave function in (\ref{eq28}) is already properly symmetrized. 
In practice, the symmetrization gives out a factor of $\sqrt{2}$ in Eqs.(\ref{eq39}) 
and (\ref{eq40}) when $j$=1. The calculation of each of these terms is formally identical to the 
case with only the elastic channel open.

\begin{table}
\caption{${\cal K}$-matrix elements are given as a function of the number
of adiabatic potentials used in the calculation ($n_A$).  }
\begin{ruledtabular}
\begin{tabular}{c c c c c}
 $n_A$         & ${\cal K}_{11}$ & ${\cal K}_{12}$  & ${\cal K}_{21}$  & ${\cal K}_{22}$  \\
\hline
2             & -2.460  & -0.650  & -0.648  & -1.411\\
3             & -2.765  & -0.821  & -0.801  & -1.496\\
4             & -2.691  & -0.775  & -0.776  & -1.468\\
6             & -2.699  & -0.781  & -0.781  & -1.471\\
8             & -2.702  & -0.783  & -0.783  & -1.471\\
10            & -2.710  & -0.787  & -0.787  & -1.473\\
14            & -2.714  & -0.790  & -0.789  & -1.474\\
18            & -2.712  & -0.791  & -0.790  & -1.474 \\

\end{tabular}
\end{ruledtabular}
\label{tab3}
\end{table}

In the calculation here we have chosen a three-body energy of $-0.7959$ mK, which 
represents an incident energy of $0.6266$ mK when channel 1 is the incoming
channel, and $0.5$ mK when channel 2 is the incoming channel. For simplicity we restrict
in this section to relative $s$-waves between the projectile and the dimer target. 
The computed result for the four terms of the ${\cal K}$-matrix are shown in table \ref{tab3}
for different values of $n_A$. 
As seen in the table, again a reduced amount of adiabatic terms permits to reach a 
reasonable convergence in the ${\cal K}$-matrix.

 From the computed ${\cal K}$-matrix we can now easily obtain the ${\cal S}$-matrix by means
of Eq.(\ref{eq27}). This leads to ${\cal S}_{11}=-0.673-0.663i$, 
${\cal S}_{12}={\cal S}_{21}=-0.285+0.162i$,
and ${\cal S}_{22}=-0.224-0.918i$. 
The square of these elements, $|{\cal S}_{ij}|^2$, 
indicates the probability for the process with incoming channel $i$ to end up in channel $j$. In 
this particular case we get $|S_{11}|^2=|S_{22}|^2=0.892$ and 
$|S_{12}|^2=|S_{21}|^2=0.108$.  

It is important to note that the matrices $A$ and $B^{2^{nd}}$ are not unique. 
A different definition of the normalization of the asymptotic states
would result into new $A$ and $B^{2^{nd}}$ matrices which 
would obviously lead to the same ${\cal K}$-matrix. In particular, $A$ and $B^{2^{nd}}$ do not
fulfill the property of being symmetric, but they lead to a 
${\cal K}$-matrix with the correct hermitian condition. Moreover, using Eq.(\ref{eq27}), 
the computed ${\cal S}$-matrix 
automatically satisfies the unitarity condition ${\cal S}^\dag {\cal S}=\mathbbm{I}$.

\section{Summary and conclusions}

In this work we have discussed the general form of the integral relations that were introduced
in \cite{bar09,kiev10}. These relations are derived from the Kohn Variational Principle and 
they permit to exploit the particularities of the adiabatic expansion method to describe scattering 
states. In particular, in \cite{bar09,kiev10} it was shown that the convergence of the computed scattering 
phase shifts in terms of the adiabatic terms included in the calculation is rather fast. The
convergence pattern results to be similar to the one of a bound state calculation. The reason for
this success is that when using the integral relations only the internal part of the wave function
is needed, and an accurate calculation of it requires a smaller amount of adiabatic terms than
when computing the wave function in the asymptotic region.

The applications given in \cite{bar09,kiev10} were limited to 
processes involving only relative $s$-waves and with only one channel open.
In this work we have explicitly derived the integral relations from the KVP in the
case of multichannel reactions and we have computed phase-shifts up to
$f$-waves. Furthermore, we have used a vectorial notation for the wave function such that all the possible
channels are simultaneously represented. With this notation the coefficients weighting the regular
and irregular part of the asymptotic wave function are $n_0\times n_0$ matrices (with $n_0$ being the
number of open channels) and each term of these two matrices is obtained from an integral relation.
Finally, the ${\cal K}$-matrix
for a given process is obtained as the product of two $n_0\times n_0$ matrices.

Although the method derived is completely general, in this work we have restricted ourselves to 
describe $1+2$
reactions with projectile energy below the breakup threshold in three outgoing particles. 
Therefore, only elastic,
inelastic, and rearrangement processes are possible.

To test the method when including relatives partial waves higher than zero, we have first 
used a toy model such that the three-body reaction is fully equivalent to a two-body process. In this
way the correct phase shift can be easily computed through a simple two-body calculation.
We have found a slow convergence of the phase shifts when extracted from the asymptotic part of
the radial wave functions. Conversely,
the rate $A^{-1}B$ converges much faster and the result stabilize with a rather small
number of adiabatic channels. The convergence
is equally fast for all the partial waves, and around 10 adiabatic terms are enough to reach
a good convergence. Furthermore, the phase shifts obtained with the two-body calculations are well
reproduced.

As the next step, we have analyzed a more physical case, in particular the \he-\dimer collision. 
Since we have considered energies below the \dimer breakup threshold, in this reaction only the elastic 
channel is open. This is a process technically analogous to the previous schematic case for which the method
has been proved to work. In fact, a similar pattern of convergence is found for the different partial 
waves included in the calculation. Inclusion of partial waves with $\ell$ up to 3 are needed to obtain
a converged cross section for the process. For $s$-waves the computed phase shift reproduces the 
one obtained with the Hyperspherical Harmonic expansion method. 

Finally, we have considered a process with two open channels. We have chosen a three-body system made
by two \he and one $^6$Li atoms, where two different dimers, \dimer and \he-$^6$Li, are possible.
We have therefore simultaneously investigated the collision between a \he atom and a \he-$^6$Li dimer, and the 
one between a $^6$Li atom and a \dimer dimer. For both reactions two possible outgoing channels (elastic
and rearrangement) are permitted. We have then used the method to obtain the $2\times 2$ ${\cal K}$-matrix.
Again, we have found a fast convergence of the four terms in ${\cal K}$. Furthermore, the computed
${\cal K}$-matrix satisfies the required hermitian condition, as well as the fact of leading to
a unitary ${\cal S}$-matrix.
 
Summarizing, we have shown that the integral relations can be easily applied to reactions
involving non-zero partial waves and with more then one channel open. Also in this case, the 
hyperspherical adiabatic expansion method is a highly efficient tool that permits to obtain
scattering wave functions. The ${\cal K}$-matrix, and therefore also the ${\cal S}$-matrix,
converges rather fast. Also, since the hyperspherical adiabatic expansion method permits to identify 
every single incoming and outgoing open channel with a single adiabatic term, the dimension of the matrices
to be computed is rather modest, typically of the same size as the number of open channels in the
reaction.

\appendix

\section{Matrix form of the Kohn Variational Principle}

This derivation is completely analogous to the derivation presented in Ref.~\cite{koh48}. 
The only difference is that, in order to represent each possible incoming channel,
we use the vectorial notation for the wave functions as introduced in the present work,
where the total wave function $\Psi$ has the form given in Eq.(\ref{eq13}).

We start by taking the matrix given by:
\begin{equation}
I\equiv \langle \Psi | \hat{\cal H}-E | \Psi \rangle
\label{eqb1}
\end{equation}
which vanishes when $\Psi $ is the exact wave function. We then introduce
a test wave function $\Psi_t=\Psi+\delta\Psi$ so that its radial wave functions
verify:
\begin{eqnarray}
&& f_{ni}^t(0)=0 \\ \nonumber 
&& f_{ni}^t(\rho\to\infty) \rightarrow 
\sqrt{k_y^{(n)}}\rho
\left( A_{in} j_{\ell_y}(k_y^{(n)}\rho) + B^t_{in} \eta_{\ell_y}(k_y^{(n)}\rho) \right).
\end{eqnarray}

We can then write $f_{ni}^t=f_{ni}+\delta f_{ni}$, where $\delta f_{ni}$ satisfies that
\begin{equation}
\delta f_{ni}(\rho) \to \sqrt{k_y^{(n)}}\rho \;
 \eta_{\ell_y}(k_y^{(n)}\rho)\;\delta B_{in}. 
\label{eqb3}
 \end{equation}

The matrix in Eq.(\ref{eqb1}), evaluated at the test wave function, is:
\begin{equation}
I_t=\delta I = \langle \Psi_t | \hat{\cal H}-E | \Psi_t \rangle
\end{equation}

Using now that the exact wave function verifies that $(\hat{\cal H}-E)|\Psi\rangle=0$,
and keeping only the first order terms, the matrix above can be written as:
\begin{equation}
\delta I = \langle \Psi | \hat{\cal H}-E | \delta\Psi \rangle
-\langle \delta \Psi | \hat{\cal H}-E | \Psi \rangle^T
\end{equation}

Using the expansion of $\Psi$ given in Eq. (\ref{eq11}) and the analytical
expression of the operator $\hat{\cal H}$ from Eq. (\ref{eq1}),
it can be seen that each matrix element of $\delta I$ takes the form:
\begin{eqnarray}
\delta I_{ij} &=& \frac{\hbar^2}{2m} \sum_n \int_0^{\infty}d\rho  
\frac{d}{d\rho}
\left( \frac{d f_{ni} }{d\rho}\delta
 f_{nj}-\frac{d (\delta f_{nj}) }{d\rho} f_{ni} \right)  \nonumber \\ 
&-& \frac{\hbar^2}{m} \sum_n \int_0^{\infty}d\rho \frac{d}{d\rho}
\left( P_{ij}(\rho) f_{ni} \delta f_{nj}\right) ,
\end{eqnarray}
where $P_{ij}(\rho)$ are the coupling terms appearing in Eq. (\ref{eq6}), and
given in Eq.(\ref{coup}), which
vanish when $\rho$ tends to zero or to infinity.  Therefore the last term
in the previous expression vanishes, and, since $f_{ni}(0)=0$, we get:
\begin{equation}
\delta I_{ij} = \frac{\hbar^2}{2m} \sum_n \left[ \frac{d f_{ni} }{d\rho}\delta
 f_{nj}-\frac{d (\delta f_{nj}) }{d\rho} f_{ni} \right]_{\rho=\infty}
\end{equation}
which, using Eq.(\ref{eq18}) and Eq.(\ref{eqb3}), leads to:
\begin{equation}
\delta I_{ij} =-\frac{\hbar^2}{2m} \sum_n A_{in} \delta B_{jn}=
-\frac{\hbar^2}{2m}\left\{ A\delta B^T \right\}_{ij},
\end{equation}
or, in a more compact way:
\begin{equation}
\delta \left(I+\frac{\hbar^2}{2m}AB^T \right)=0.
\end{equation}

Since for the exact wave function $\Psi$ we have that $I=0$, we finally get:
\begin{equation}
\frac{\hbar^2}{2m}AB^T=I_t+\frac{\hbar^2}{2m}AB^T_t,
\end{equation}
which becomes a variational principle for $AB^T$. Therefore, given a test wave
function $\Psi_t$, we obtain a second order correction for  $AB^T$ as:
\begin{equation}
A(B^{2^{nd}})^T=AB^T+\frac{2m}{\hbar^2}
 \langle \Psi_t | \hat{\cal H}-E | \Psi_t \rangle
\end{equation}

If we now multiply from the left by $A^{-1}$ and from the right by $(A^{-1})^T$,
and make use of the fact that the ${\cal K}$-matrix is symmetric,
i.e. $A^{-1} B= B^T (A^{-1})^T$, we then finally get the expression given in Eq.(\ref{eq41}).

\section{Calculation of the integrals in $A$ and $B$}

In this appendix we give details of the calculation of the integrals $A$ and $B$,
in particular when using two-body potentials projecting on partial waves.
To this aim let us start from the
general expression for $A$ and $B$ in Eq.\refeq{eq47}, and write the ${\cal L}$
operator in its explicit form
\begin{eqnarray}
& &{\cal L}= \frac{2m}{\hbar^2}(\hat{{\cal H}}-E)=\frac{2m}{\hbar^2}
\Big(- \frac{\hbar^2}{2m}\bigtriangledown^2_{\bm{y_1}} \\ \nonumber
& & -\frac{\hbar^2}{2m}\bigtriangledown^2_{\bm{x_1}} 
 +\hat{V}_1(x_1)+\hat{V}_2(x_2)+\hat{V}_3(x_3)
-E_d-E_0 \Big)
\end{eqnarray}
where $E_d$ is the binding energy of the dimer and $E_0$ is the
incident energy of the projectile. The Jacobi coordinates $(\bm{x}_1,\bm{y}_1)$
are defined such that $\bm{x}_1$ connects the two particles in the dimer (particles
2 and 3). The coordinates $x_2$ and $x_3$ are related to the distances between particles
1 and 3, and between particles 1 and 2, respectively.

Using $F$ and $G$ as defined in \refeq{eq28} we can rewrite Eq.\refeq{eq47} as:
\begin{eqnarray}
\label{intab}
B^{2^{nd}}&=&\frac{2m}{\hbar^2}<\Psi|\hat{V}_2(x_2)+\hat{V}_3(x_3)|F(\bm{x}_1,\bm{y}_1)>  \\ \nonumber
A&=& -\frac{2m}{\hbar^2}<\Psi|\hat{V}_2(x_2)+\hat{V}_3(x_3)|{\widetilde G}(\bm{x}_1,\bm{y}_1)> 
+ I_{\bigtriangledown}
\end{eqnarray}
where
\begin{equation}
I_{\bigtriangledown} =-
<\Psi|\bigtriangledown^2_{\bm{y_1}}-k_{y_1}^2 |{\widetilde G}(x_1,y_1)>,
\label{intgam}
\end{equation}
and where $\widetilde G$ refers to the regularized function (\ref{eq28b}).

If we call $B^{2^{nd}}=I_B$ and $A=I_A+I_{\bigtriangledown}$, we have that,
after substitution of Eq.(\ref{eq28}), the integrals in (\ref{intab}) can be written 
as well as:
\begin{eqnarray}
I_{A,B}=\frac{2m}{\hbar^2}\sum_{i=2,3}\int d\rho \rho^5 d\Omega_i 
\Psi(\rho,\Omega_i) \hat{V}_i(x_i)\label{eqb4} \\ \nonumber
g_{A,B}^{\ell_{y_1}}(y_1,k_{y_1}) 
\left[ \psi_{\ell_{x_1}}(\bm{x_1})\otimes Y_{\ell_{y_1}}(\Omega_{y_1})\right]^{LM_L}
\end{eqnarray}
where
\begin{eqnarray}
g_B^{\ell_y}(y,k_y)&=&j_{\ell_y}(k_y y) \\ \nonumber
g_A^{\ell_y}(y,k_y)&=&-\eta_{\ell_y}(k_y y) \left(1-e^{\gamma y}\right)^{\ell_y+1},
\end{eqnarray}
and where for simplicity in the notation we have assumed that the particles have zero spin.
The corresponding expressions in this appendix for particles with spin will follow immediately by
coupling the orbital part in the expressions above to the corresponding spin part.

If the potential operator is given as a sum of projectors on partial waves, we have that:
\begin{equation}
\hat{V}_i(x_i)=\sum_{\ell_{x_i} m_{\ell_{x_i}}} V_{\ell_{x_i}}(x_i) |\ell_{x_i} m_{\ell_{x_i}}
\ket \bra \ell_{x_i} m_{\ell_{x_i}}|,
\end{equation}
where $V_{\ell_{x_i}}$
represents the interaction between particles $j$ and $k$ when they are in a relative 
partial wave with angular momentum $\ell_{x_i}$.  

If we also consider Eq.(\ref{eq3}) and expand the angular functions
$\Phi_n(\rho,\Omega)$ in terms of the hyperspherical harmonics 
($\Phi_n(\rho,\Omega)^{LM} = \sum_{K \ell_x \ell_y} C_{K \ell_x \ell_y L}^{(n)}(\rho) 
\mathcal{Y}^{K L M}_{\ell_x \ell_y}(\Omega)$), we can then obtain the following
expression for the potential operator acting over the three-body wave function:
\begin{eqnarray}
\langle \Psi |\hat{V}(x)=\frac{1}{\rho^{5/2}}\sum_n f_n(\rho)
\sum_{K \ell_x \ell_y}\sum_{\tilde{K} \tilde{\ell}_x \tilde{\ell}_y}
C_{\tilde{K} \tilde{\ell}_x \tilde{\ell}_y L}^{(n)} \label{eqb7} \\ \nonumber
\Big\bra \mathcal{Y}^{\tilde{K} L M}_{\tilde{\ell}_x\tilde{\ell}_y}|V_{\ell_x}(x)|
\mathcal{Y}^{K L M}_{\ell_x \ell_y}\Big\ket 
\Big\bra\mathcal{Y}^{K L M}_{\ell_x \ell_y}| .
\end{eqnarray}

Expanding now the hyperspherical harmonics in terms of the Jacobi polynomials 
$P_\nu^{\ell_x+1/2, \ell_y+1/2}$ ($K=2\nu+\ell_x+\ell_y$) with normalization coefficients
$N^{\ell_x \ell_y}_{K}$ (see \cite{nie01} for details), we have that the
the integrals $I_A$ and $I_B$ can then be explicitly written as:
\begin{eqnarray}
I_{A,B}& &=\sum_{i=2,3} \int d\rho \rho^5 (\sin \alpha_i)^2(\cos \alpha_i)^2 
d\alpha_i
\frac{1}{\rho^{5/2}}   \label{eqb8}\\ \nonumber 
& &
\sum_n f_n(\rho) \sum_{K \ell_x \ell_y}\sum_{\tilde{K} \tilde{\ell}_x
\tilde{\ell}_y}
C_{K \ell_x \ell_y L}^{(n)} N^{\tilde{\ell}_x \tilde{\ell}_y}_{\tilde{K}} (\sin
\alpha_i)^{\tilde{\ell}_x}(\cos \alpha_i)^{\tilde{\ell}_y}  \\ \nonumber
& & 
P_\nu^{\tilde{\ell}_x+1/2,
\tilde{\ell}_y+1/2}(2\cos \alpha_i) 
\Big\bra \mathcal{Y}^{\tilde{K} L M}_{\tilde{\ell}_x\tilde{\ell}_y}|V_{\ell_x}(x_i)|
\mathcal{Y}^{K L M}_{\ell_x \ell_y}\Big\ket  \\ \nonumber
& & 
R_{i1}^{\tilde{\ell}_{x_i} \tilde{\ell}_{y_i},\ell_{x_1} \ell_{y_1}} 
\left[ \phi_d^{\ell_{x_1}}(x_1)g_{A,B}^{\ell_{y_1}}(k_{y_1},y_1)\right]
\end{eqnarray}
where $\phi_d^{\ell_{x_1}}$ is the radial part of the dimer wave function 
$\psi_{\ell_{x_1}}(\bm{x_1})$.

It is important to note that in Eqs.(\ref{intab}) and (\ref{eqb4}) the potential operators 
and the functions $F$ and $\widetilde{G}$ are written in a different Jacobi set.
Therefore, when computing the integrals one has to rotate the whole integrand into the same Jacobi set. 
This is made in the expression above by the function $R_{ij}$, which is a rotation function defined as:
\begin{eqnarray}
& & R_{ij}^{\ell_{x_i} \ell_{y_i},\ell_{x_j} \ell_{y_j}} \left[ W_{\ell_{x_j} \ell_{y_j}}(x_j,y_j)
\right] =  \\ \nonumber 
& & \int d\Omega_{x_i} d\Omega_{y_i}  
\left[ Y_{\ell_{x_i}}^*(\Omega_{x_i})\otimes Y_{\ell_{y_i}}^*(\Omega_{y_i})
\right]^{LM_L} \\ \nonumber
& & W_{\ell_{x_j} \ell_{y_j}}(x_j,y_j)\left[ Y_{\ell_{x_j}}(\Omega_{x_j})
\otimes Y_{\ell_{y_j}}(\Omega_{y_j})\right]^{LM_L} \\ \nonumber
\end{eqnarray}
which rotates any function $W_{\ell_{x_j} \ell_{y_j}}(x_j,y_j)$ written
in terms of the coordinates and angular momenta defined in the Jacobi set $j$ into
the coordinates and angular momenta corresponding to the Jacobi set $i$.

As already mentioned, $V_\ell(x)$ is the total two-body interaction when the two particles are
in a relative partial wave with angular momentum $\ell_x$. In general, for particles with spin, the 
partial waves are identified by the quantum numbers $\{\ell_x,s_x,j_x\}$, where $s_x$ is the coupling
of the spins of the two particles, which in turn couples to $\ell_x$ to give the total two-body
angular momentum $j_x$. In this case the matrix element 
$\Big\bra \mathcal{Y}^{\tilde{K} L M}_{\tilde{\ell}_x\tilde{\ell}_y}|V_{\ell_x}(x)|
\mathcal{Y}^{K L M}_{\ell_x \ell_y}\Big\ket$ in Eqs.(\ref{eqb7}) or (\ref{eqb8}) has to
be replaced by:
\begin{equation}
\Big\bra \left[\mathcal{Y}^{\tilde{K} L}_{\tilde{\ell}_x\tilde{\ell}_y}\otimes \chi_{s_x,s_y}^S
\right]^{JM}\left|V_{\ell_x s_x j_x}(x)\right|
\left[\mathcal{Y}^{K L }_{\ell_x \ell_y}\otimes \chi_{s_x,s_y}^S\right]^{JM}\Big\ket,
\label{eqb10}
\end{equation}
where $s_y$ is the spin of the third particle, $\chi_{s_x,s_y}^S$ is three-body spin wave function
and $J,M$ are the total three-body angular momentum and its projection. In general, the partial
wave two-body potential $V_{\ell_x s_x j_x}(x)$ could consist in a sum of central, 
spin-orbit, spin-spin and tensor potentials. Therefore, in this case, calculation of the
matrix element in (\ref{eqb10}) implies calculation of the matrix element of the corresponding
spin-spin, spin-orbit, and tensor operators. In the simplest case with only a central potential 
and particles with zero spin the matrix elements of the potential operator reduce to:
\begin{equation}
\Big \bra \mathcal{Y}^{\tilde{K} L M}_{\tilde{\ell}_x\tilde{\ell}_y}|V_{\ell_x}(x)|
\mathcal{Y}^{K L M}_{\ell_x \ell_y}\Big\ket=V_{\ell_x}(x) \delta_{\ell_x,\tilde{\ell}_x}
\delta_{\ell_y,\tilde{\ell}_y}\delta_{K,\tilde{K}},
\end{equation}
simplifying the expression (\ref{eqb8}).

\acknowledgments
This work was partly supported by funds provided by DGI of MEC (Spain)
under contract No.  FIS2008-01301. One of us (C.R.R.) acknowledges
support by a predoctoral I3P grant from CSIC and the European
Social Fund.


\begin{thebibliography}{99}
 
\bibitem{gloeckle94} W. Gl\"ockle {\sl et al.}, Phys. Rep. {\bf 274}, 107 (1996).
\bibitem{deltuva07} A. Deltuva and A.C. Fonseca, Phys. Rev. C {\bf 75}, 014005 (2007).
\bibitem{kiev01} A. Kievsky, M. Viviani and S. Rosati, Phys. Rev. C {\bf 64}, 024002 (2001).
\bibitem{kiev08} A. Kievsky, S. Rosati, M. Viviani, L.E. Marcucci, L. Girlanda, 
J. Phys. G {\bf 35}, 063101 (2008).
\bibitem{benchmark1} A. Kievsky {\sl et al.}, Phys. Rev. C {\bf 58}, 3085 (1998).
\bibitem{benchmark2} R. Lazauskas, J. Carbonell, A.C. Fonseca, M. Viviani, A. Kievsky,
 and S. Rosati, Phys. Rev. C {\bf 71}, 034004 (2005).
\bibitem{motovilov} E.A. Kolganova, A.K. Motovilov and W. Sandhas, Phys. Part. Nuc. {\bf 40}, 206 (2009).
\bibitem{bar01} P. Barletta and A. Kievsky, Phys. Rev. A {\bf 64}, 042514 (2001).
\bibitem{nie01} E. Nielsen, D.V. Fedorov, A.S. Jensen, and E. Garrido,
Phys. Rep. {\bf 347}, 373 (2001).
\bibitem{blume00} D. Blume, Ch.H. Greene and B.D. Esry, J. Chem. Phys. {\bf 113}, 2145 (2000).
\bibitem{esry2008} H. Suno and B.D. Esry, Phys. Rev. A {\bf 78}, 062701 (2008).
\bibitem{greene2010} Ch.H. Greene, Phys. Today {\bf 63}, 40 (2010).
\bibitem{bar09b} P. Barletta and A. Kievsky, Few-Body Syst. {\bf 45}, 25 (2009).
\bibitem{bar09} P. Barletta, C. Romero-Redondo, A. Kievsky, M. Viviani, E. Garrido,
Phys. Rev. Lett {\bf 103}, 090402 (2009).
\bibitem{har67} F.E. Harris, Phys. Rev. Lett. {\bf 19}, 173 (1967).
\bibitem{hol72} A.R. Holt and B. Santoso, J. Phys. B {\bf 5}, 497 (1972).
\bibitem{wang09} Y. Wang and B.D. Esry, Phys. Rev. Lett {\bf 102}, 133201 (2009).
\bibitem{kiev10} A. Kievsky, M. Viviani, P. Barletta, C. Romero-Redondo, E. Garrido,
Phys. Rev. C {\bf 81}, 034002 (2010).
\bibitem{nil98} E. Nielsen, D. V. Fedorov and A.S. Jensen, J. Phys. B {\bf 31}, 4085 (1998).
\bibitem{azi91} Aziz R.A. and Slaman M.J., J. Chem. Phys. {\bf 94}, 8047 (1991).
\bibitem{incao08} J.P. D'Incao, B.D. Esry and Ch.H. Greene, Phys. Rev. A {\bf 77}, 052709 (2008).
\bibitem{stecher09} J. von Stecher and Ch.H. Greene, Phys. Rev. A {\bf 80}, 022504 (2009).
\bibitem{sun08} H. Suno and B.D. Esry, Phys. Rev A {\bf 78}, 062701 (2008).
\bibitem{rou03} V. Roudnev, Chem. Phys. Lett. {\bf 367}, 95 (2003).
\bibitem{mot01} A.K. Motovilov, W. Sandhas, S.A. Sofianos, and
E. A. Kolganova, Eur. Phys. J. D {\bf 13}, 33 (2001).
\bibitem{kol04} E.A. Kolganova, A.K. Motovilov, and W. Sandhas, Phys. Rev. A {\bf 70}, 052711 (2004).
\bibitem{suno09} H. Suno and B.D. Esry, Phys. Rev A {\bf 80}, 062702 (2009).
\bibitem{kle99} U. Kleinekath\"{o}fer, M. Lewerenz and M. Mladenovi\'{c},
Phys. Rev. Lett. {\bf 83}, 4717 (1999).
\bibitem{koh48} W. Kohn, Phys. Rev. {\bf 74}, 1763 (1948).

\end{thebibliography}
\end{document}